\documentclass[onecolumn]{aa} 
\usepackage{graphicx}
\usepackage{txfonts}
\usepackage{hyperref}
\hypersetup{colorlinks=true, allcolors=blue}
\usepackage{siunitx}
\begin{document} 

\title{A\&A community survey on the future of scientific publishing}
\subtitle{Credibility over speed, fairness over profit, human judgment over automation}
\authorrunning{Alves et al.}
\author{Jo\~ao Alves\inst{1}, Ar\=unas Ku\v{c}inskas\inst{2}, Charlotte Van Rooyen\inst{3}, Marc Audard\inst{4}, Pierre-Alain Duc\inst{5}, David Elbaz\inst{6}, Thierry Forveille\inst{7}, Laszlo L. Kiss\inst{8,9}, Tiago Pereira\inst{10,11}, Eva Villaver\inst{12} 
}
\mail{joao.alves@univie.ac.at} \institute{%
  University  of Vienna, Department of Astrophysics, T\"urkenschanzstrasse 17, 1180 Vienna, Austria 
  \and Institute of Theoretical Physics and Astronomy, Vilnius University, Saul\.{e}tekio al. 3, Vilnius, LT-10257, Lithuania
  \and EDP Sciences, 17, avenue du Hoggar, P.A. de Courtab{\oe}uf, B.P. 112, F-91944 Les Ulis Cedex, France
  \and Department of Astronomy, University of Geneva, Chemin Pegasi 51, 1290 Versoix, Switzerland
  \and Universit\'e de Strasbourg, CNRS, Observatoire astronomique de Strasbourg, UMR 7550, F-67000 Strasbourg, France
  \and Université Paris-Saclay, Université Paris Cité, CEA, CNRS, AIM, 91191 Gif-sur-Yvette, France
  \and Université Grenoble Alpes, CNRS, IPAG, F-38000, Grenoble, France
  \and Konkoly Observatory, HUN-REN Research Centre for Astronomy and Earth Sciences, H-1121 Budapest, Konkoly Th.M. 15-17, Hungary
  \and ELTE E\"otv\"os Lor\'and University, Institute of Physics and Astronomy, Budapest, Hungary
  \and Rosseland Centre for Solar Physics, University of Oslo, P.O. Box 1029 Blindern, NO--0315 Oslo, Norway
  \and Institute of Theoretical Astrophysics, University of Oslo, P.O. Box 1029 Blindern, NO--0315 Oslo, Norway
  \and Instituto de Astrofísica de Canarias, 38200 La Laguna, Tenerife, Spain
  } 
\date{Received: April 2026 ; Accepted  }

\abstract 
{
Scientific publishing is undergoing major change, driven by a shift toward open access (OA), the rise of artificial intelligence (AI), and growing demands for transparency, reproducibility, and equity. At the same time, rapid growth in article output strains editors and reviewers and means that metrics and speed can eclipse quality and rigor. To better understand how the community is responding, 
Astronomy \& Astrophysics (A\&A) commissioned the {A\&A Survey on Trends and Challenges in Scientific Publishing}, which documents community opinion on journal choice, peer review, OA, research evaluation, and the role of AI, with the goal of informing future editorial policies and the wider conversation on sustainable, ethical, and equitable scientific communication. Distributed online in May 2025 to \SI{28787} A\&A authors and co-authors, the 13-question survey drew \SI{2944} responses from 69 countries by its closing date. Quantitative answers were analyzed through frequency distributions, and roughly \SI{3000} free-text comments were grouped into recurring themes. The responses were clear. Journal quality and reputation are the most decisive factors in deciding where to publish, followed by cost. The principal worry about peer review is reviewer expertise and fairness rather than speed. Citation counts are still an important consideration, but many respondents want broader, more qualitative measures of impact. The majority of them prefer public or institutional funding for OA, and views on AI are polarized, with widespread acceptance of administrative and language assistance but firm opposition to autonomous decision-making or content generation. Integrity, credibility, and fairness are common themes in every section of the responses. Overall, the survey portrays an engaged community that values quality over speed, fairness over profit, and human oversight over automation, providing A\&A with clear insight into community preference and a solid framework for shaping future policies on OA, peer review, and the responsible integration of AI.
} 

   \keywords{Community Survey}
   \maketitle
%
\section{Introduction}

Journals are shifting from subscription-based access, where content is available only to paying institutions or individuals, to open access (OA) models that make research freely available, as in the case of Astronomy \& Astrophysics (A\&A). At the same time, metrics to evaluate careers in most countries, as well as AI \citep{van2023ai}, are reshaping how researchers communicate and assess scientific work. The pressure to move beyond narrow metrics has prompted community responses such as the San Francisco Declaration on Research Assessment \citep{DORA2013}, the Leiden Manifesto \citep{hicks2015bibliometrics}, and the Agreement on Reforming Research Assessment \citep{CoARA2022}.

As one of the major journals in the field, A\&A plays a leading role in this changing environment. Established as a nonprofit community-owned publication, {``a Journal for astronomers by astronomers,''} A\&A has always aimed to balance scientific excellence with accessibility and fairness. Because publishing practices evolve rapidly, the journal needs to have regular dialogue with its authors. In May 2025, A\&A launched the {Survey on Trends and Challenges in Scientific Publishing}, targeting its broad base of authors and co-authors. The survey aimed to gauge attitudes toward journal selection, peer review, research evaluation, OA, and the integration of AI into publishing workflows.

The response was impressive: \SI{2944} participants from 69 countries provided quantitative ratings and extensive free-text comments. The resulting dataset offers one of the most comprehensive contemporary snapshots of researchers' perspectives on scientific publishing in astronomy. Beyond statistical trends, the richness of the almost \SI{3000} free-text comments reveals the values that guide the community: a commitment to integrity, fairness, and responsible research communication. A preliminary analysis of the survey data was presented at the Annual Meeting of the European Astronomical Society in Cork in June 2025. This Editorial presents and interprets the main results of the survey, drawing on both the quantitative responses to the survey questions and the qualitative insights provided through free-form comments by individual A\&A authors. This lets us report not only what the community thinks, but the reasoning behind it.

\begin{figure}
    \centering
    \includegraphics[width=0.9\linewidth]{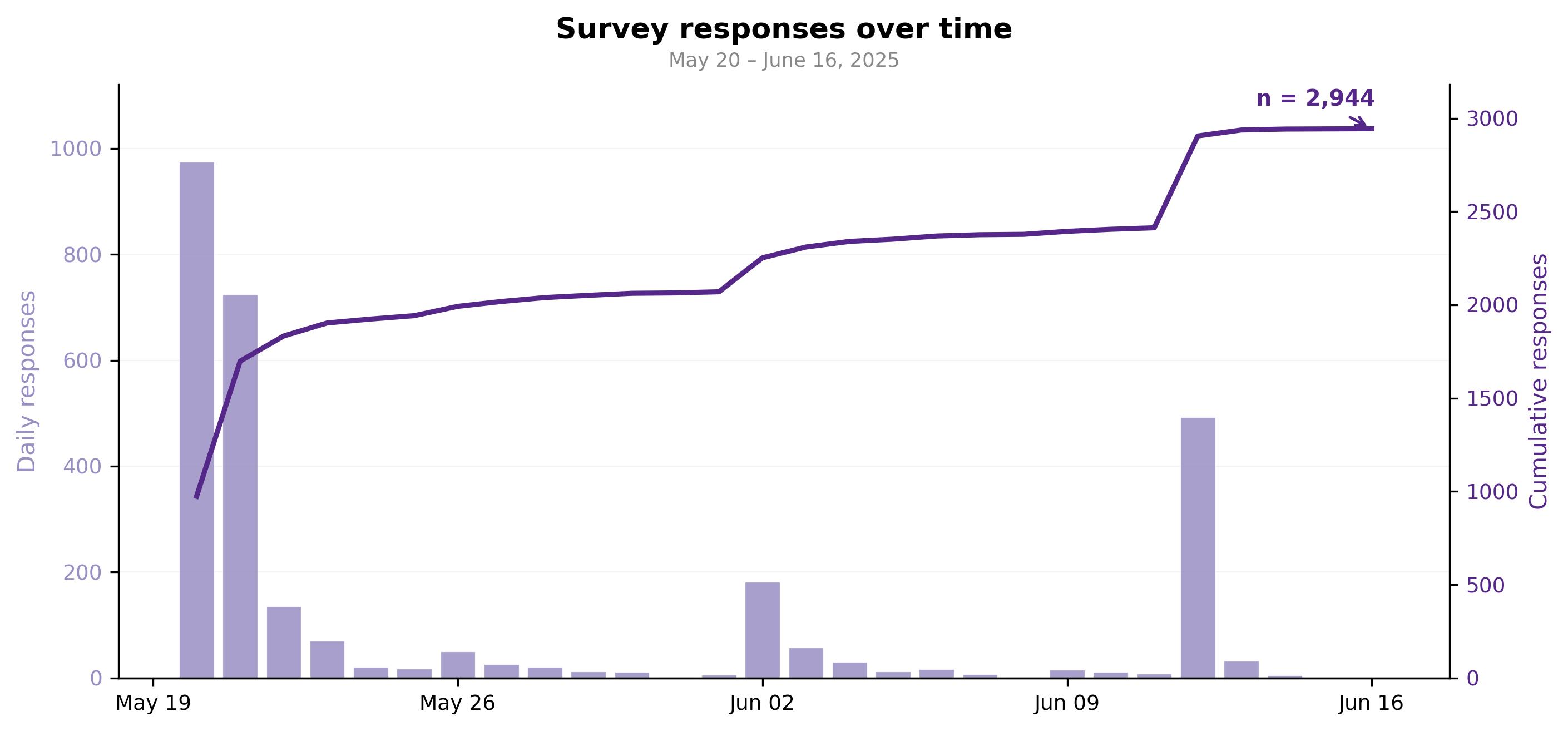}
    \caption{Cumulative number of responses over time. The survey opened on 20 May 2025 and closed on 16 June 2025. Reminder emails at the end of May and just before  the survey closed on 16 June 2025 produced visible increases in participation.}
    \label{fig:responses}
\end{figure}

\section{Survey design and participation} 
\subsection{Objectives and structure}

The {A\&A Survey on Trends and Challenges in Scientific Publishing} was designed to gather broad, quantitative, and qualitative feedback from the journal's author community on key aspects of scientific publishing. The questionnaire addressed 13 topics grouped into five thematic areas:

\begin{enumerate}
\item Journal selection criteria;
\item The peer-review process;
\item Evaluation of research impact;
\item OA and funding models; 
\item The role of AI in publishing.
\end{enumerate}

Each thematic block consisted of multiple-choice questions using a five-point scale (from ``least important'' = 1 to ``most important'' = 5) and open text fields for comments. These qualitative responses were later coded by theme to complement the statistical analysis. The {A\&A Survey on Trends and Challenges in Scientific Publishing} aimed to capture the collective experience and expectations of the community rather than test predefined hypotheses. In this sense, it was exploratory and intended to guide A\&A's editorial and strategic decisions on policy, communication, and future directions.

\begin{figure}
    \centering
    \includegraphics[width=0.9\linewidth]{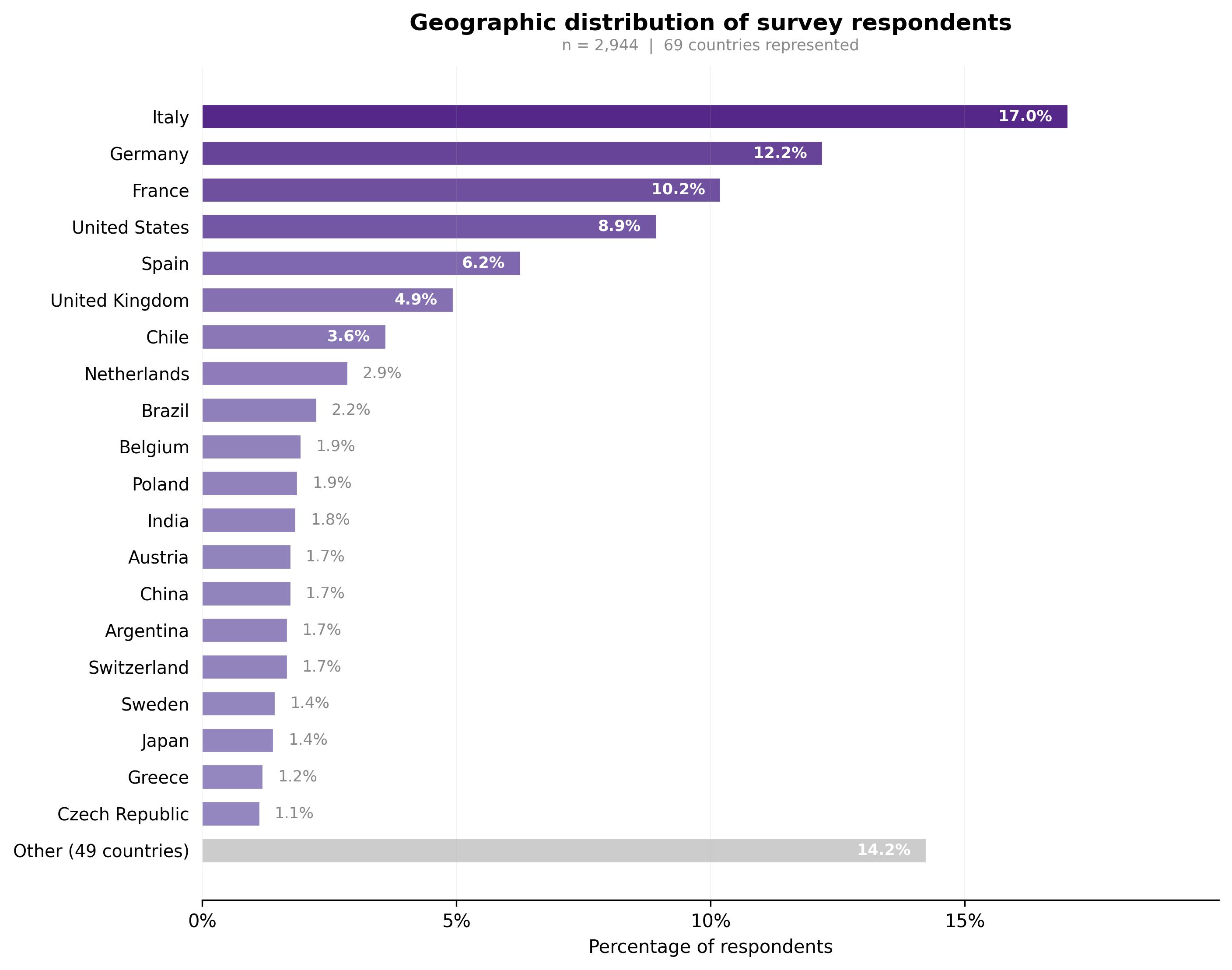}
    \caption{Geographic distribution of respondents to the {A\&A Survey on Trends and Challenges in Scientific Publishing}. The survey reached participants in 69 countries, with the highest representation in Europe but substantial participation worldwide.}
    \label{fig:countries}
\end{figure}

The questionnaire was distributed on 20 May 2025 to \SI{28787} A\&A authors and co-authors from the previous three years, using EDP Sciences’ email distribution platform. The survey closed on 16 June 2025, at which time \SI{2944} responses had been recorded. To put this level of participation into perspective and considering that the A\&A community is mostly from A\&A sponsor countries,  this number is comparable to roughly half the membership of the European Astronomical Society and a quarter of that of the International Astronomical Union, providing a useful sense of scale relative to the active astronomy community. The substantial response rate suggests genuine interest across the community in the development and future of A\&A and its role in scientific publishing.

The response timeline (see Fig.~\ref{fig:responses}) shows a strong initial surge after the announcement, with distinct secondary peaks after the reminder emails were sent at the end of May and just before the closing date. This pattern indicates that the initial communication channels reached a large portion of the target audience effectively, while reminders increased participation by prompting those who had not yet responded. The respondents represent 69 countries (see Fig.~\ref{fig:countries}), which confirms the international character of A\&A. The highest participation came from Europe, followed by North America and Asia, with approximately a third (29\%) of all responses from non-sponsor countries.

\subsection{Career stages and demographics}

The respondents represent all stages of an academic career, from PhD students and postdocs to senior researchers and emeritus scientists. The breakdown between junior and senior researchers is roughly balanced (PhD students and postdocs, 42\%; mid-career and senior scientists, 55\%). A smaller fraction of them are retired or emeritus scientists. 
Because both senior and junior researchers are well represented, the survey captures established views alongside those of newer authors. Separating the responses into junior and senior groups does not change the main results of the survey. The largest difference between these two groups appears in the answers to{ ``What role could AI play in the editorial process?,''} with the junior group being more receptive to AI usage.\ However, even for that question, the difference is modest.

\subsection{Data analysis and limitations}

The responses were collected anonymously. Quantitative data were analyzed using standard statistical summaries (means, medians, and frequency distributions). The almost \SI{3000} anonymous free-text comments (about \SI{60000} words) were reviewed manually and with the support of large language models (LLMs). The open text comments were not used for model training and were summarized into recurring themes such as costs of publishing, bias in peer review, OA funding, and AI ethics.
The main limitations of the survey stem from self-selection bias and our mailing-list reach: (1) active authors are likely overrepresented relative to occasional contributors; (2) regional participation may correlate with the journal's established readership; and (3) opinions expressed in comments, while informative and appreciated, may overrepresent more vocal viewpoints. Nevertheless, the scale and diversity of responses provide a reliable overview of the prevailing attitudes within the astronomy publishing community. The dataset thus constitutes a robust empirical foundation for analyzing community expectations. The following sections present the results thematically, combining statistical trends with the most frequently expressed qualitative opinions.




\begin{center}

    \includegraphics[width=0.9\linewidth]{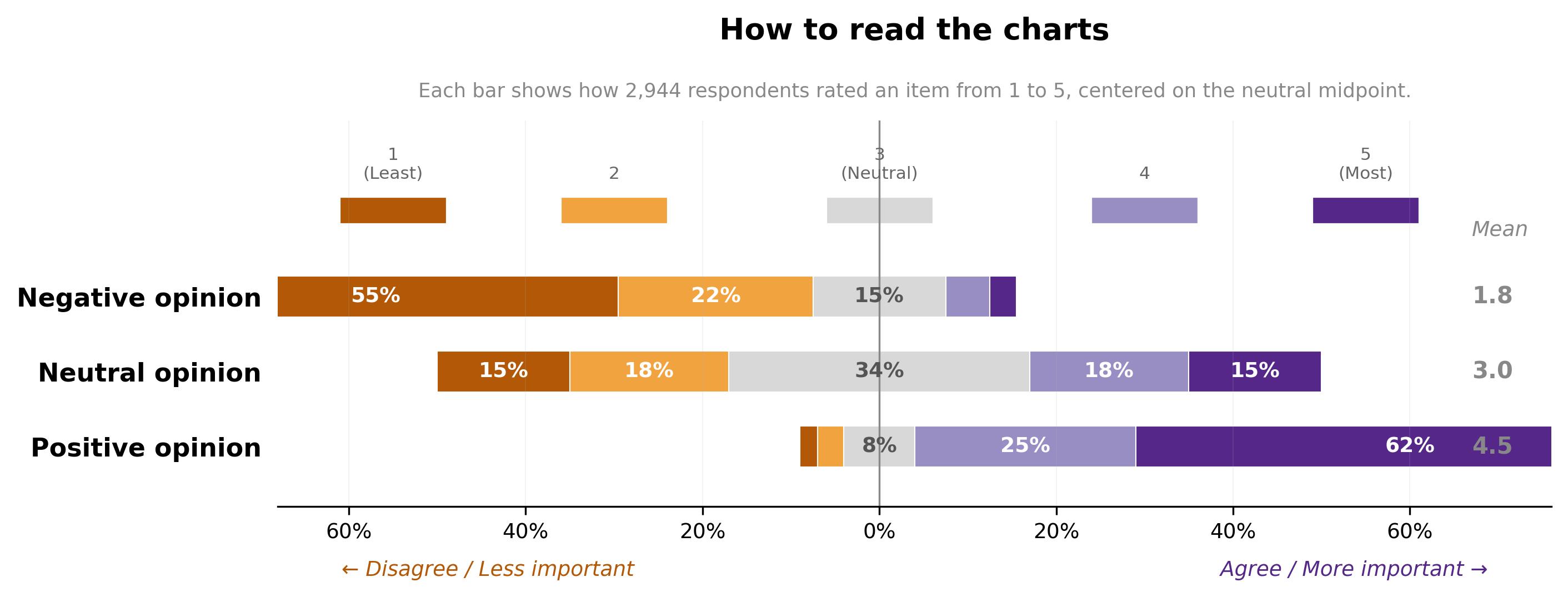}
    \label{fig:explainer}
        
\end{center}
In the following figures (as indicated in Fig. 3), the x-axis shows the percentage of respondents who gave each rating (1 through 5), displayed using a diverging bar chart with the center as the neutral rating (3).  The negative side (left of zero) represents all ratings 1, 2, and half of 3. The positive side (right of zero) represents the other half of rating 3 plus all of 4 and 5. The percentage labels on the x-axis (20\%, 40\%, etc.) represent the cumulative percentage of respondents going outward from the center in each direction. So, if a bar extends to ``40\%'' on the right, that means roughly 40\% of respondents gave that item a rating of 4 or 5 (plus half of the 3s). Finally, the zero line represents the center of the neutral opinion. When a bar extends much further to the right than to the left, the community leans positive on that item. When the bars are roughly symmetric, opinion is split. The mean column on the right gives the simple arithmetic average of all 1--5 ratings for that item as a single summary number. 


\section{Choosing a journal}

\begin{figure}
    \centering
    \includegraphics[width=0.9\linewidth]{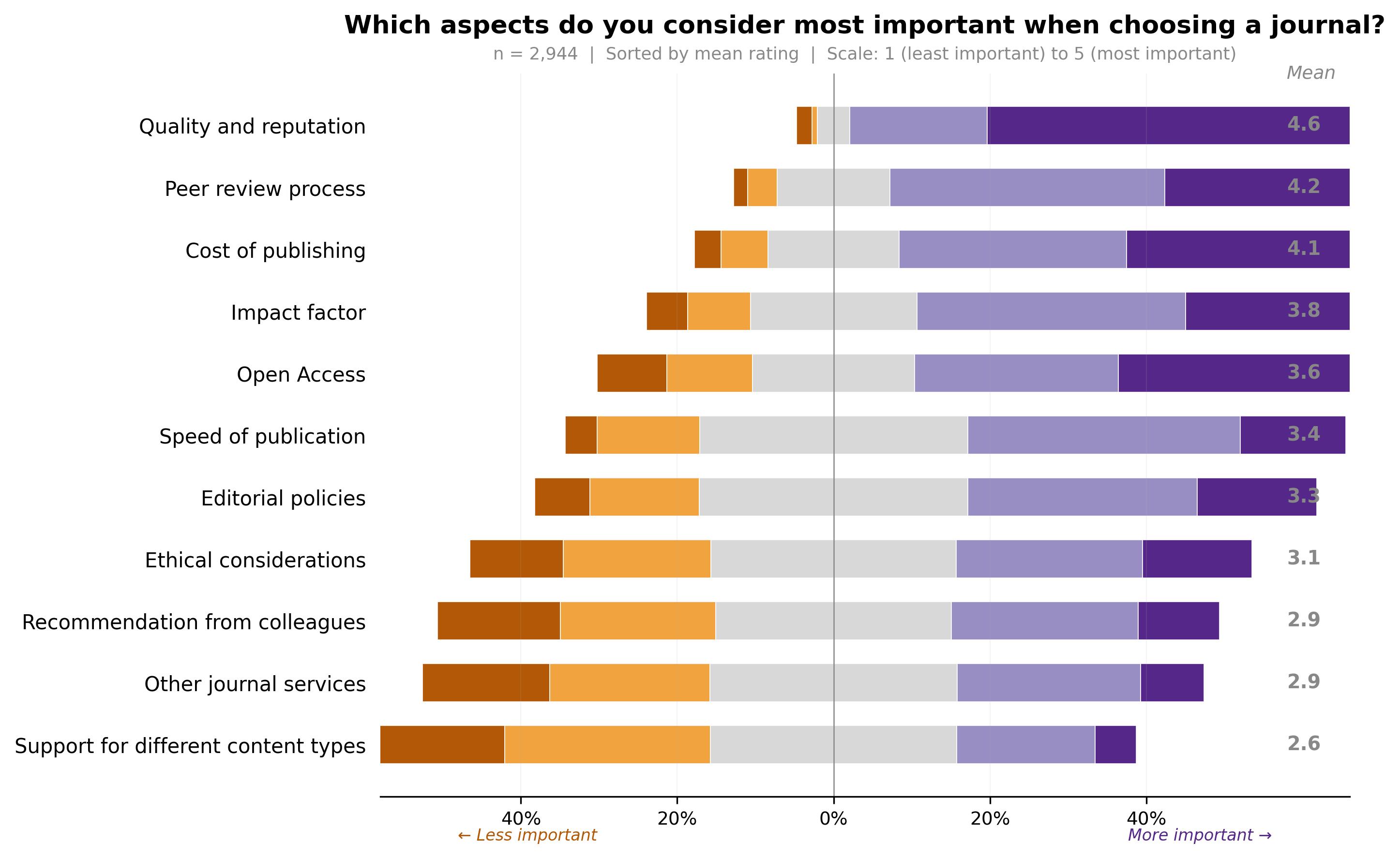}
    \caption{Relative importance of factors influencing journal choice. Journal quality and reputation dominate decision-making, followed by publication cost. Speed, impact metrics, and journal scope are of secondary concern.}
    \label{fig:choosingjournal}
\end{figure}

When choosing where to publish, the respondents care more about the quality and reputation of the journal (see Fig.~\ref{fig:choosingjournal}, consistent with long-established patterns in researcher behavior, and more recent analyses of publishing decisions, \citealt{Xu2023}). They see a journal's scientific credibility, its editorial standards, and the fairness and rigor of its peer review as the real measure of value. Cost is the following most important consideration, with many highlighting the financial strain of article processing charges in OA journals. Other factors such as publication speed, impact factor, and journal scope matter less to them. Overall, the responses show that astronomers still place trust and reliability well above metrics or fast turnaround times. Several respondents also noted that publication costs should not come at the expense of editorial quality or integrity, a theme explored further in Sect.~\ref{sec:openaccess}.

\begin{figure}
    \centering
    \includegraphics[width=0.9\linewidth]{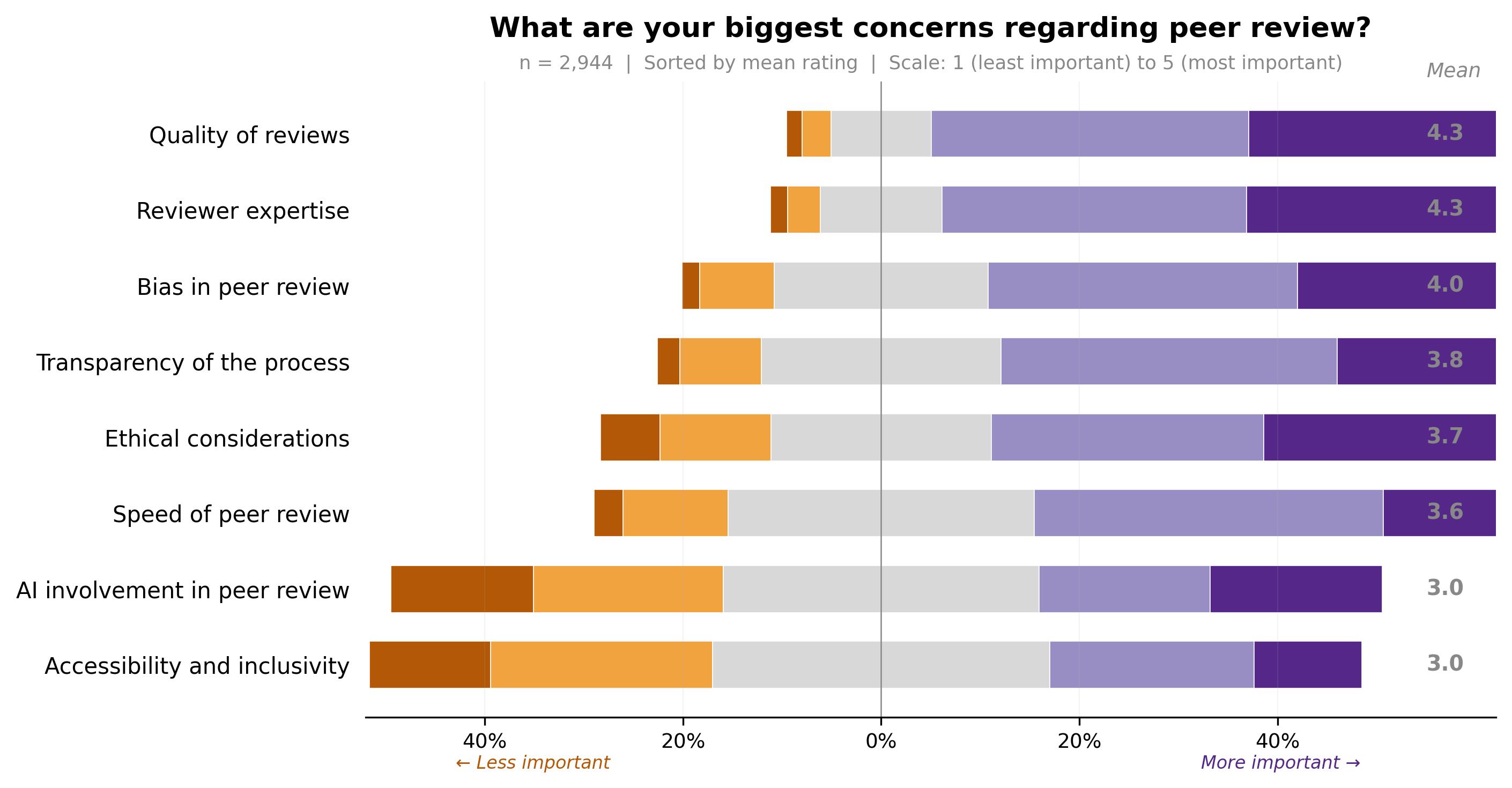}
    \caption{Main concerns about peer review reported by respondents. Reviewer expertise and bias in peer review stand out as the top issues, followed by the quality of feedback and review speed. The results show that the community values review quality more than quick turnaround times.}
    \label{fig:concerns}
\end{figure}

\section{Peer review: Strengths and weaknesses}

Peer review remains central to how research is evaluated and trusted, and respondents share strong views about its strengths and weaknesses. Although they value efficiency and transparency, most are far more concerned with the quality and expertise of the reviewers (see Fig.~\ref{fig:concerns}). Many noted that the integrity of the editorial process depends on reviewers who truly understand the subject and apply fair, consistent, and constructive standards when evaluating a paper.

The free-text responses reiterated the three main concerns identified earlier: the expertise of the reviewer, fairness, and the quality of the feedback. Many researchers described situations in which referees did not have sufficient knowledge of the topic, leading to miscommunication or unsuitable feedback. Others mentioned inconsistencies between reviewers, limited editorial oversight, or a lack of accountability when reviews were perceived to be superficial or dismissive. Fairness was another recurring theme, and several respondents noted perceived bias linked to institution, region, gender, or personal connections between authors and reviewers. Views vary on implementation: some respondents support the wider use of double-blind review to reduce bias, while others favor open review, where reviewer identities and reports are disclosed, to make the process more transparent and accountable \citep[for a multidisciplinary review of peer-review models and innovations, see][]{lee2013bias,Tennant2017,RossHellauer2017}.

Many respondents see the lack of constructive feedback as one of the biggest weaknesses in peer review. They feel that too often reviews focus on rejection or correction instead of helping authors improve their work and that vague or overly negative comments do not provide constructive guidance. The participants strongly support clearer editorial standards and better training or recognition for reviewers. Suggestions include formally acknowledging the review activity, as well as offering fee discounts or even small financial rewards to encourage more thoughtful and engaged reviews.

Although the speed of review matters, most respondents do not see it as the main issue. They are willing to accept moderate turnaround times if it means that the review was careful and fair. Many noted that preprints already provide rapid dissemination, so journals should instead focus on thorough evaluation and the certification of results. Several also call for clearer editorial decisions and better communication with the authors throughout the review process.

\begin{figure}
    \centering
    \includegraphics[width=0.9\linewidth]{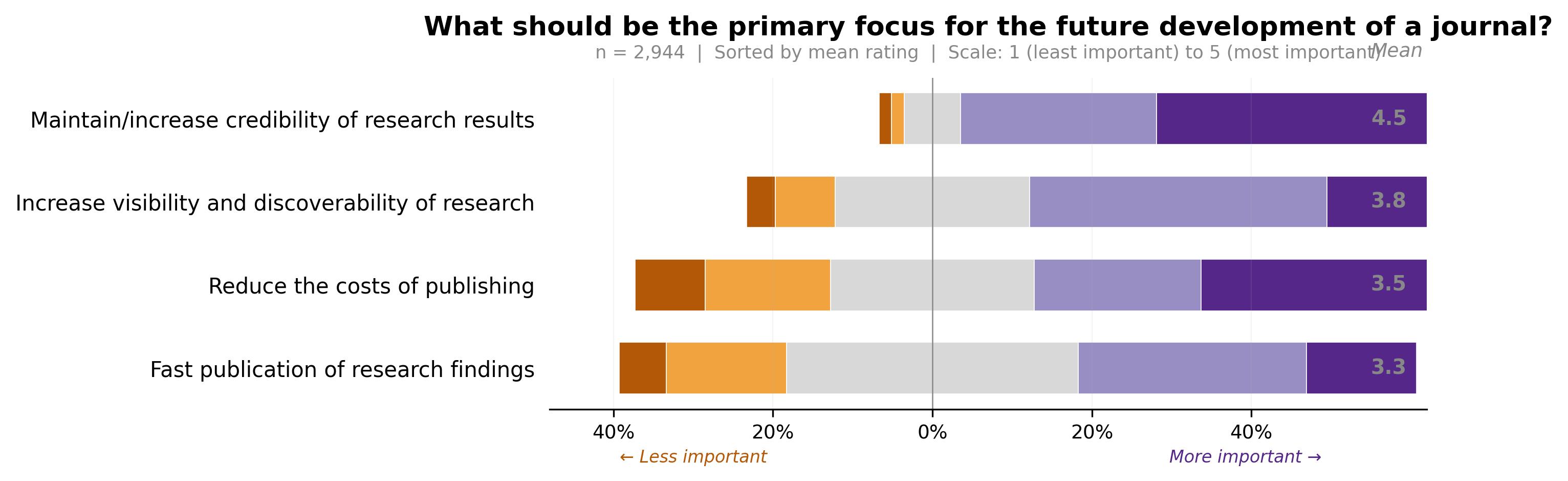}
    \caption{Priorities for the future development of journals. Respondents overwhelmingly emphasized credibility as a primary focus for the future development of a journal, ahead of visibility, speed, and cost reduction.}
    \label{fig:future}
\end{figure}

\section{The future of journals}

When asked about the future of scientific journals, respondents have one clear priority: maintaining and strengthening the credibility of published research (see Fig.~\ref{fig:future}). This concern outweighed all others, including publication speed and visibility. The responses show a community deeply committed to the integrity of the scientific record and cautious about practices that could weaken trust in the literature. Many participants also raised concerns about publication costs and the equity of access, themes addressed in detail in Sect.~\ref{sec:openaccess}.


Respondents place a strong emphasis on reproducibility and data integrity. 
Many urge journals to require authors to share data, code, and analysis workflows, expectations that closely correspond to the Findable, Accessible, Interoperable, and Reusable (FAIR) principles of \citealt{Wilkinson2016}, although the framework itself was not explicitly referenced in the survey. Some suggest creating dedicated “reproducibility editors” or adding verification steps to the editorial process. The participants also linked credibility to avoiding ``salami publishing,'' where the results are split into as many articles as possible. Instead, they prefer fewer publications that present more complete and coherent studies.

Some respondents encourage journals to expand their visibility and outreach through multimedia, plain-language summaries, or AI tools that improve discoverability. As preprints already provide rapid dissemination, respondents see the journal's future role primarily in certification and long-term curation of the scientific record.

\begin{figure}
    \centering
    \includegraphics[width=0.9\linewidth]{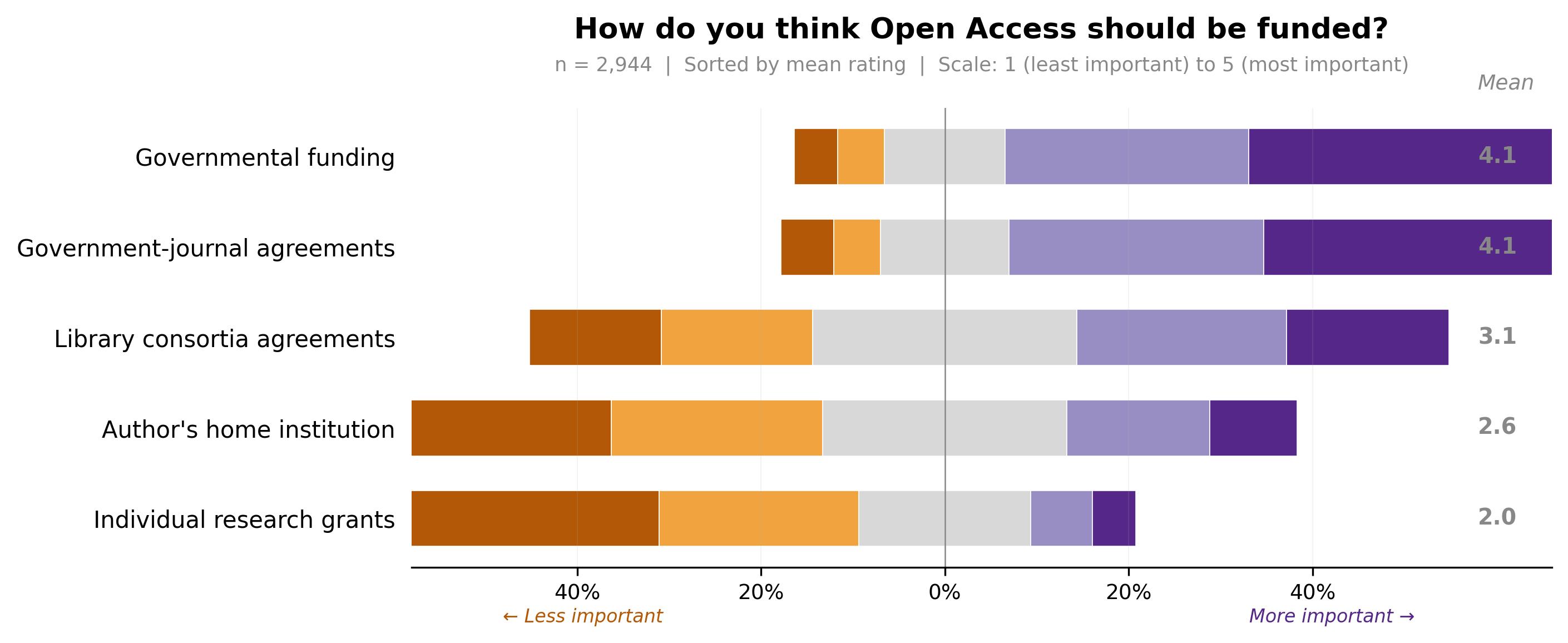}
    \caption{Perceived importance of different OA-funding models. Governmental funding and government-journal agreements are clearly preferred, while library consortia occupy an intermediate position. Funding via authors' institutions and individual research grants are considered the least favored options.}
    \label{fig:funding}
\end{figure}

\begin{figure}
    \centering
    \includegraphics[width=0.9\linewidth]{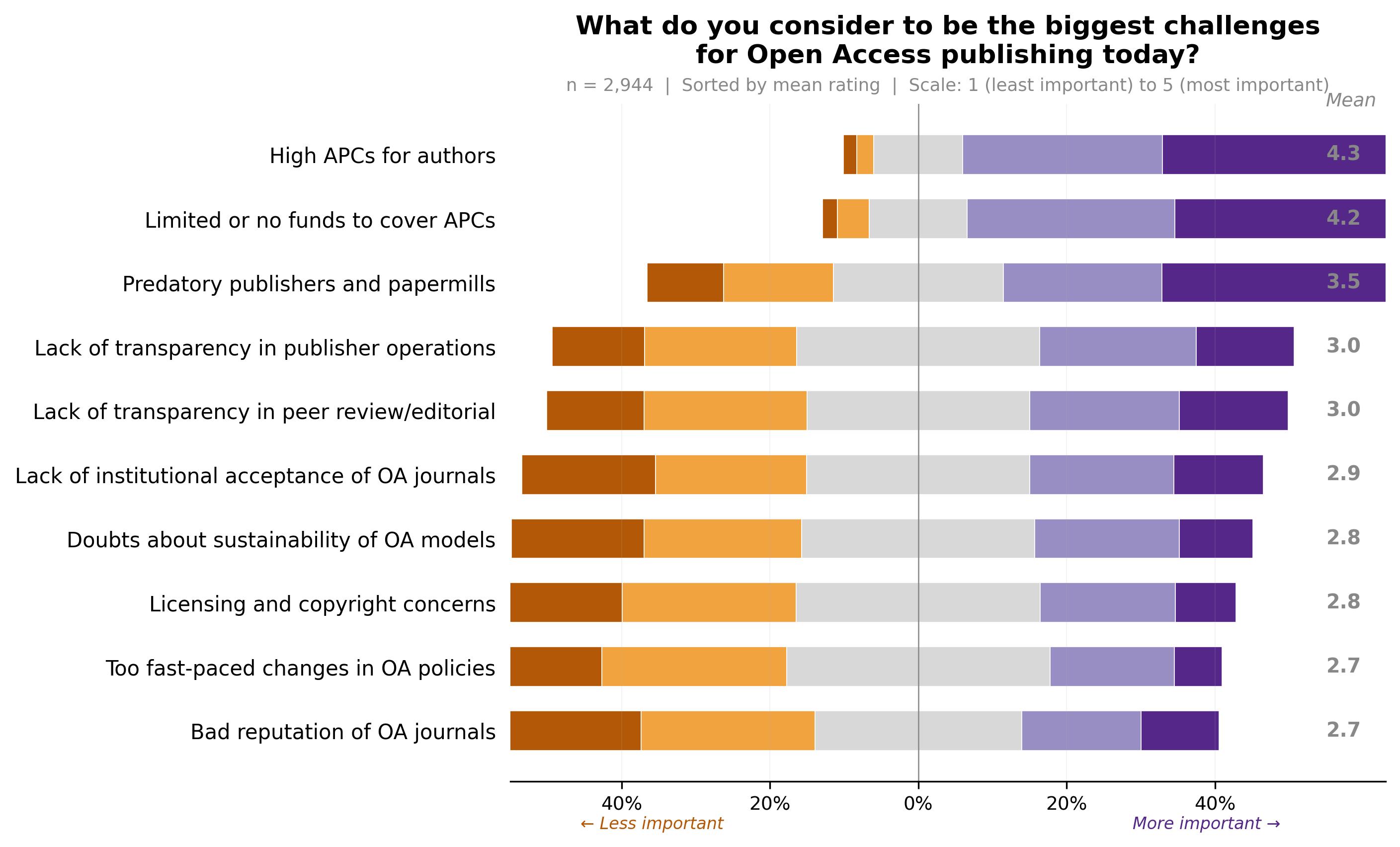}
    \caption{Perceived challenges for OA publishing. High article processing charges (APCs) and limited funding to cover them emerge as the dominant concerns. Issues related to predatory publishing follow, while transparency, institutional acceptance, sustainability, and policy pace are viewed as moderate challenges.
    }
    \label{fig:challenges}
\end{figure}

\section{Open access: Attitudes, challenges, and funding models}\label{sec:openaccess}

The survey indicates that OA publishing is at the center of today's debates on the future of scientific communication and shows that astronomers strongly support the idea of OA, the free public availability of research. However, they are highly critical of models in which authors pay, which dominate current practice  (see Fig.~\ref{fig:funding}). Their comments draw a clear line between the moral value of openness and the economic systems used to sustain it.

Most of the respondents said that the high article processing charges (APCs) had stopped them at least once from submitting to a journal they preferred. Many described these fees as unreasonable, unfair, or unsustainable. Researchers from institutions or countries with limited funding expressed particular frustration, noting that APCs widen global inequalities and make publishing harder for early-career scientists. Even those at well-funded institutions said that publication budgets are already stretched too thin to handle rising costs. It is important to note that while these concerns reflect widespread frustration with APC-based models common in commercial publishing, A\&A does not charge APCs. Instead, A\&A operates as a community-supported journal with a different funding structure.

Another recurring argument concerns the role of preprint servers, especially arXiv, which already ensure free global access to nearly all astronomical research. Many therefore question the necessity of paying APCs for OA journal versions that add little accessibility beyond what preprints provide. For these respondents, the unique value of journals lies in the peer-review process and long-term archiving, not in access to articles per se. Several propose that journals focus on certification and quality assurance, while open dissemination could remain with repositories.



The survey also asked about the perceived benefits and challenges of OA publishing. Respondents most often cite greater accessibility, easier exchange between disciplines, and greater visibility for publicly funded research as key advantages. The main challenges are high APCs (see Fig.~\ref{fig:challenges}), unequal access to publication funds, the rise of profit-driven publishing models, and the difficulty in distinguishing reputable OA journals from predatory ones. In terms of funding, respondents agree that OA should be supported collectively rather than by individual authors and that publishing should be possible regardless of personal or institutional resources. Several noted that existing agreements between journals and institutions made publishing easier, while others said they were unable to publish at all without such arrangements. Waivers are widely appreciated and considered essential for authors in developing countries. A smaller but notable group advocate for fully nonprofit or community-led publishing models, reflecting astronomy's long-standing culture of collaboration.

Astronomy \& Astrophysics's funding model already reflects many of these community values. As a nonprofit community-owned journal, A\&A does not operate on a standard APC model and it is made OA through collective funding. Publishing costs are supported through a combination of revenue streams. Institutional sponsorships from national and international organizations contribute approximately 25\% of operating expenses, while the remaining income is derived from library subscriptions under the Subscribe to Open (S2O) model \citep{Crow2020}, together with modest page charges for authors in non-sponsoring countries and for excess page lengths.

This mixed model aims to balance sustainability with equitable access, while avoiding reliance on full APC funding. However, maintaining A\&A in OA under S2O requires sustained commitment and support from the community.

\section{Artificial intelligence in publishing}

\begin{figure}
    \centering
    \includegraphics[width=0.9\linewidth]{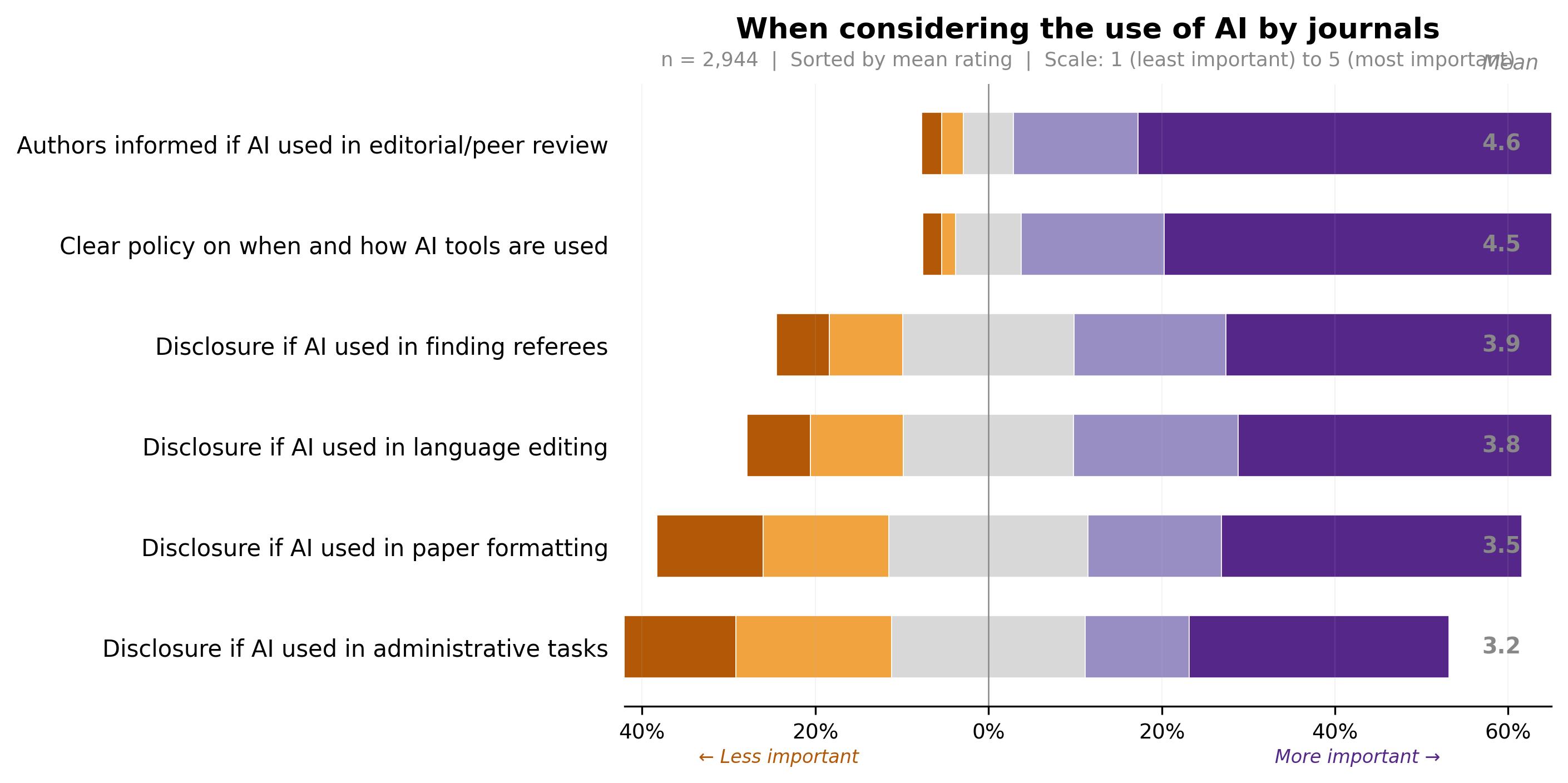}
    \caption{Expectations regarding the use of AI by journals. Transparency measures are strongly supported, particularly informing authors when AI is used in editorial or peer-review processes and establishing clear policies on AI use. Disclosure of AI use in referee selection, language editing, and formatting is also considered important, while disclosure for administrative tasks is viewed as less critical.
    }
    \label{fig:ai_journals}
\end{figure}

\begin{figure}
    \centering
    \includegraphics[width=0.9\linewidth]{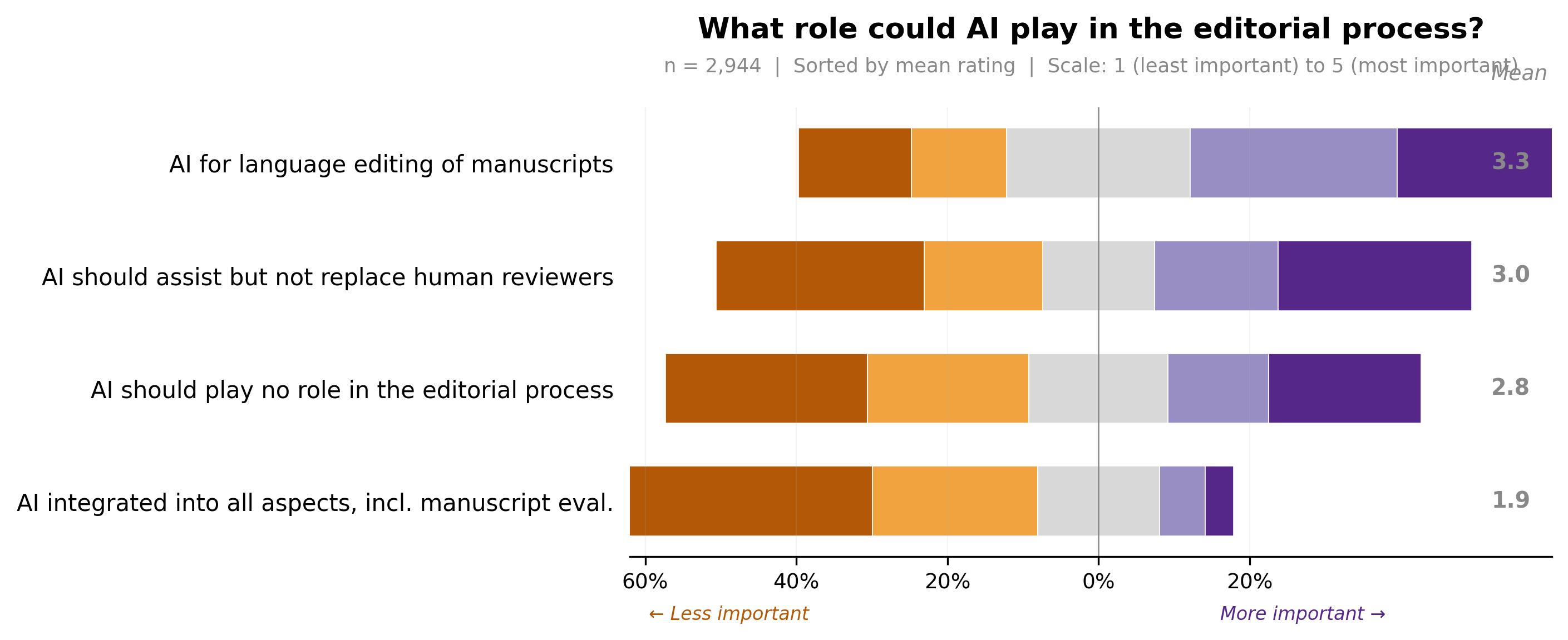}
    \caption{AI is most supported for language editing of manuscripts in the editorial process, followed by a role assisting but not replacing human reviewers. There is limited support for excluding AI entirely, and there is strong opposition to the full integration of AI into all aspects of the editorial process.
    }
    \label{fig:ai_editorial}
\end{figure}

\begin{figure}
    \centering
    \includegraphics[width=0.9\linewidth]{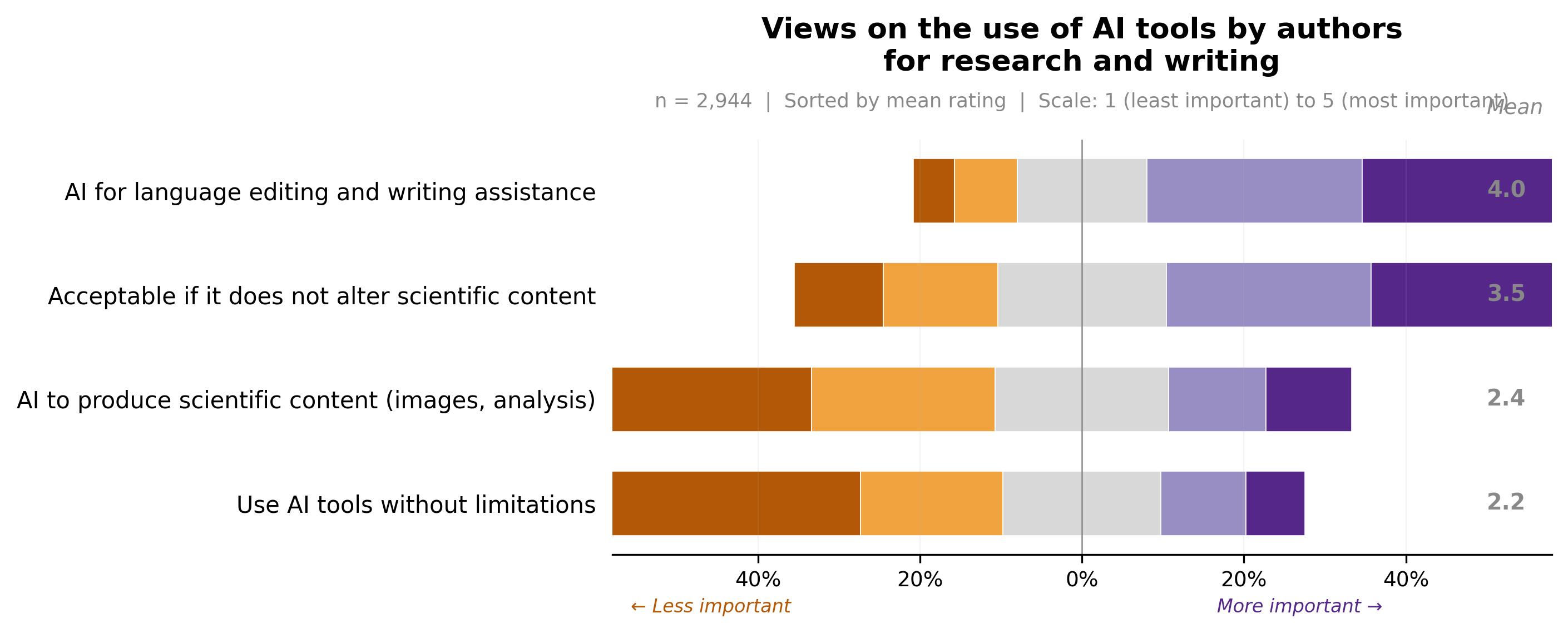}
    \caption{AI is strongly supported for language editing and writing assistance for authors, and it is considered acceptable if it does not alter scientific content. There is limited support for using AI to generate scientific content or without restrictions.
    }
    \label{fig:ai_authors}
\end{figure}



\subsection{Artificial intelligence in editorial workflows and journal policies}

Many respondents view AI-assisted reviewer suggestions positively, provided that editors maintain oversight (see Fig.~\ref{fig:ai_journals}). The respondents feel that such tools can improve efficiency and reduce workload without compromising the scientific evaluation. Most of the respondents support using AI for routine editorial or administrative tasks (see Fig.~\ref{fig:ai_editorial}). They see value in tools that help to screen manuscripts for completeness or plagiarism, check figures and references, and improve language quality and grammar. At the same time, the survey reveals a nuanced view of the role of AI in the editorial process. A substantial fraction of respondents favor limiting or excluding AI from the editorial process, and there is clear opposition to the full integration of AI into all aspects of manuscript evaluation. In general, the results indicate that the community accepts AI as a supporting tool but not as a decision-making agent.

The expectations regarding the journal's use of AI emphasize transparency (Fig.~\ref{fig:ai_journals}). The most important measures are to inform authors when AI is used in editorial or peer-review processes and to establish explicit policies on when and how AI tools are applied. Disclosure of AI use in specific tasks, such as referee selection, language editing, or formatting, is also considered important, though with a slightly lower priority. In contrast, disclosure of AI use in purely administrative tasks is viewed as less critical. Overall, these results point to a consistent position: AI can play a useful role in improving efficiency and supporting editorial workflows, but it must remain under human oversight and guided by a transparent policy.

\subsection{Artificial intelligence use by authors}
When it came to the use of AI by authors, the respondents again support it as a tool rather than a creator (see Fig.~\ref{fig:ai_authors}). They view applications such as machine learning for data analysis, code generation or debugging, figure preparation, and language editing, especially for non-native English speakers, as appropriate. In contrast, most reject generative text or image models that produce scientific content without human validation. Many call for mandatory disclosure of AI use in manuscripts, including which tools were used and for what purpose. Several also suggest that journals treat AI tools as any other software, requiring proper citation and ensuring reproducibility.

A smaller but vocal minority argue for tighter restrictions or outright bans on AI-generated text, warning that such tools can introduce hidden errors, fabricated references, or overly uniform writing styles. Others recognize that AI will inevitably become part of research, but stress that authors must remain fully responsible for their content. Many agree that education, clear journal policies, and well-defined ethical guidelines would work better than prohibitions that are difficult to enforce.

\subsection{Broader ethical reflections}
Beyond practical concerns, many respondents raised deeper ethical and cultural questions about the growing use of AI in science. They caution against the overreliance on automated systems, the loss of writing and critical thinking skills, and the concentration of power in a few corporations that control large AI models. Most agree that transparency, reproducibility, and fairness should guide how AI is integrated into scientific work. Some express cautious optimism, suggesting that AI could improve accessibility and efficiency if used responsibly, but none see it as a replacement for human reasoning or judgment.

\section{Conclusions and outlook}

The {A\&A Survey on Trends and Challenges in Scientific Publishing} offers a comprehensive view of how astronomers see the current transformation of scholarly communication. A common thread runs throughout: credibility has to remain the foundation of publishing, regardless of technology or a particular business model.

Three priorities emerge from the survey. First, A\&A must maintain its editorial rigor and peer-review quality, the attributes respondents value most when choosing a journal. Second, the journal's nonprofit, community-funded model positions it well in a landscape where the approach in which authors pay is widely considered unsustainable, though sustaining it requires ongoing institutional commitment in a challenging funding environment. Third, the community has given a clear mandate on AI: it is useful as a support tool, unacceptable as a decision-maker, and acceptable only with a transparent policy and human accountability.

The survey points to a clear vision for A\&A and for publishing in astronomy: rigorous, fair, open, and grounded in human judgment. We are deeply grateful to the community for this remarkable level of participation. Without the thoughtful participation of nearly \SI{3000} astronomers, we would not have this invaluable guidance for shaping A\&A's future.

A\&A is committed to translating these insights into a concrete strategy. The scale and depth of participation itself confirm that this community expects and values an active role in shaping how its science is published. Building on our long-standing identity as a {"journal for astronomers by astronomers" that is nonprofit and} community-owned, we will organize town hall discussions to continue this dialogue with the community. We are developing comprehensive policies on AI that reflect the principles articulated in this survey, particularly in terms of transparency and human accountability. More broadly, we are using these results to guide our strategic planning, ensuring that A\&A remains responsive to the evolving needs of astronomical research, while preserving the values of credibility, equity, and trust that the community has clearly affirmed in a landscape where the forces shaping scholarly communication are becoming increasingly complex and interdependent \citep{STMTrends2030}.

Periodic surveys such as this one will help maintain an ongoing dialogue between A\&A and the authors. In a time of rapid change, the results remind us that the foundation of scholarly publishing is not technology but trust and that trust must be earned, maintained, and shared by the entire community.

\begin{acknowledgements}
The authors acknowledge the help and support from the EDP team and the stimulating discussions with Uta Grothkopf. Large language models were used to support language editing, to summarize the large volume of free-text responses (about 60000 words), and to assist in coding the final figures. 
\end{acknowledgements}

\bibliographystyle{aa} 
\bibliography{references}






\end{document}